# GaAs doped by self-assembled molecular monolayers

Zhengfang Fan[1], Yumeng Liu[1], Yizuo Wang[1], Shuwen Guo[1,2*], Li He[3*], Yaping Dan[1*]

[1]University of Michigan – Shanghai Jiao Tong University Joint Institute, Shanghai Jiao Tong University, Shanghai, 20240 China

[2]School of Energy and Materials, Shanghai Polytechnic University, Shanghai 201209, China

[3]State Key Laboratory of Superlattices and Microstructures, Institute of Semiconductors, Chinese Academy of Science, Beijing, 100083, China

Email: shuwen.guo@sjtu.edu.cn, heli2018@semi.ac.cn, yaping.dan@sjtu.edu.cn

Abstract

Self-assembled molecular monolayer doping remains as a research focus for its nature of being conformal, nondestructive, and self-limiting. Herein, we demonstrate a sulfur monolayer doping in GaAs, facilitated by $(NH_4)_2S_x$ solution. The Van der Pauw technique, secondary-ion mass spectroscopy, and low-temperature Hall effect measurements show that the sulfur dopants concentration and electron activation rate are $4\times10^{20}$ cm$^{-3}$ and 77.6%, respectively. The donor energy level of sulfur-doped GaAs is located 68 meV below the conduction band. Based on this process, a p-n junction was successfully fabricated on highly doped p-type GaAs substrate.

Introduction

Self-assembled molecular monolayer (SAMM) doping has a wide range of applications in nano-electronic semiconductor devices, complementary metal oxide semiconductor (CMOS) field effect transistors, and Fin Field-Effect Transistor (FinFET)[1,2]. This technique enables precise control over dopant profiles, which is crucial for achieving optimal device performance. It has been successfully demonstrated for a wide range of n- and p-type dopants on silicon[3,4](Si) and germanium[5] (Ge) surfaces, demonstrating its versatility and effectiveness. Recently, the method has been extensively applied to III-V semiconductors with exceptional transport characteristics, which make them particularly suitable for applications in optoelectronics, high-speed nanoelectronics[6,7], and solar energy harvesting[8], where precise control over dopant profiles is crucial for achieving optimal device performance.

For III-V substrates, most studies have mainly used sulfur-passivation protocols as a precursor to n-type doping[9-12]. The treatment of GaAs with $(NH_4)_2S_x$ solution results in a sulfur monolayer on its surface. This monolayer is particularly advantageous as it is free of dangling bonds, thus preventing the adsorption of foreign atoms[10,13,14]. This characteristic is crucial for maintaining a clean and stable surface, which is vital for achieving precise doping profiles. Furthermore, the chemisorbed S atoms can be utilized as a source of S doping for the GaAs substrate. Hedieh et.al[15] reported a stable passivated GaAs surface with the help of saturated ammonium sulfide solution and the absence of re-oxidation for several days post-exposure to air. Mattson et.al[16] investigated the thermal evolution of sulfur-passivated GaAs (100) without any additional capping layers, using a combination of in situ characterization techniques and found sulfur subsurface diffusion is characterized by a (2×1) reconstruction, arising from dimerization of the terminal S atoms. Kenneth R. Kort et.al[17] studied the dopant activation in sulfur monolayer doped $In_{0.53}Ga_{0.47}As$ samples, and found a yield carrier densities as high as $9.5×10^{18}$ cm$^{-3}$. However, seldom studies have investigated the electrical activation rate of dopants after sulfur monolayer doping in GaAs.

In this work, we demonstrate a SAMM doping technique by treating GaAs in $(NH4)_2S$ solution (sulfur can be grafted onto GaAs) with a rapid thermal annealing (RTA) process. The-SAMM doping is~50 nm and the peak concentration is as high as ~$4×10^{20}$ cm$^{-3}$, in which the electron activation rate is 77.6%. In addition, the ideality factor of the fabricated p-n junction in p-type GaAs is 1.26, indicating a good junction quality.

Experimental details

The SAMM-doping process using sulfur precursors on the GaAs substrate is shown in figure 1. Undoped GaAs (100) wafers were chemically cleaned with acetone, isopropanol and deionized water. For sulfur monolayer formation, a saturated ammonium sulfide solution was prepared by adding 0.27g excess sulfur to a 20 mL $(NH4)_2S$ solution, which was stirred for 2 h to obtain a proper dissolution of sulfur excess. Prior to sulfur monolayer formation, the native oxide was removed by immersion in a 3:1 solution of deionized water/HCl(37%) for 5 min followed by a 3 min deionized water rinse. Subsequently, the GaAs wafers were dipped into a saturated $(NH_4)_2S$ solution for 4 h at room temperature[16]. After S monolayer formation on the GaAs surface, a 200 nm thick $SiO_2$ capping layer was sputtered onto the samples, followed by annealing at different temperatures to drive S atoms into GaAs. Finally, a layer of 5 nm germanium and 100 nm gold electrode were deposited on the surface of the samples by a thermal evaporation system. After the evaporation, the metals were annealed at 400 °C for 30 s by RTA under an atmospheric pressure of $N_2$ to obtain ohmic contact[18].

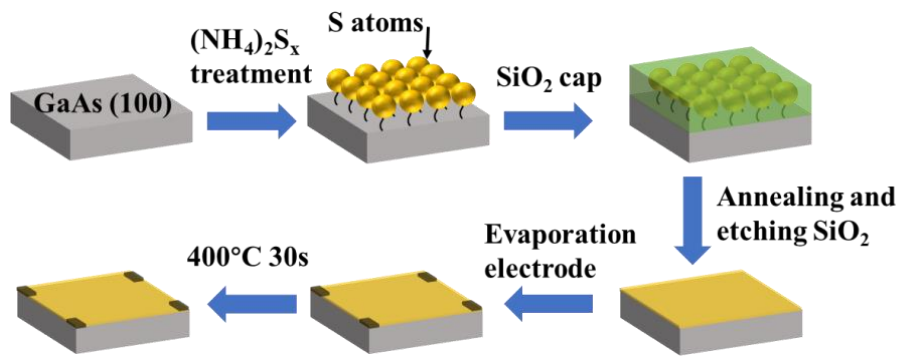

Figure 1. Schematic of SAMM-doping process using sulfur precursors on the GaAs substrate.

Results and discussion

X-ray photoelectron spectroscopy (XPS) was performed for the surface-functionalized sample as shown in figure 2, which shows the XPS spectrum of Ga 2p, As 3d, and S 2p core levels of sulfur-passivated GaAs surfaces. In Figure 2a, a shoulder can be clearly observed on the higher bonding energy side of the Ga 2p peak. After curve-fitting, a second peak with a chemical shift of 0.53 eV is observed, which is consistent with the formation of Ga-S bonding. As shown in figure 2b, the peak at 41.2eV is related to Ga-As bond, and As elemental is located at a binding energy of 42eV, which is in good agreement with the literature[15]. Due to the overlap of Ga 3s and As 2p plasmon loss, the S 2p XPS spectra is more complex. Figure 2c clearly indicates several different chemical states, including Ga 3s level (binding energy at 160eV), S 2p level (at 162eV) and a plasmon satellite from the intense As 2p[11,13]. Collectively, these XPS data indicate that sulfur atoms occupy the dangling bonds at the surface of GaAs, indicating that sulfur atoms have been grafted on the GaAs surfaces.

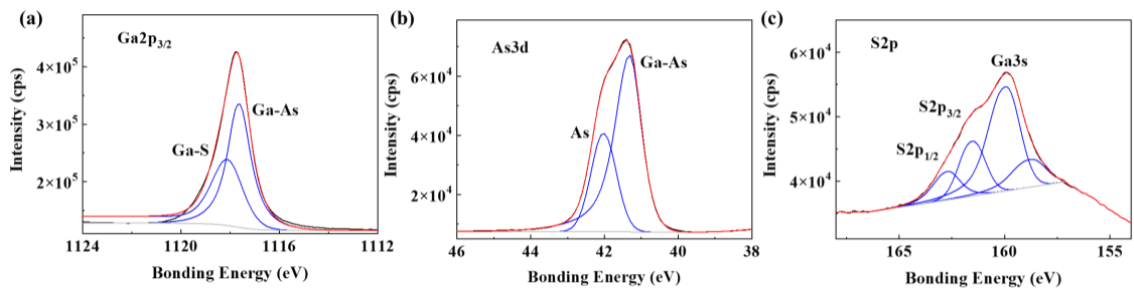

Figure 2. (a) Ga 2p, (b) As 3d, and (c) S 2p core-level XPS spectrum of S-passivated GaAs surfaces.

After successfully grafting sulfur atoms onto the surfaces of GaAs, a 200 nm thick $SiO_2$ capping layer was sputtered onto the samples to act as a protective barrier layer during RTA. By varying the annealing temperature, different doping concentration levels were attained, enabling us to investigate the impact of sulfur doping on the electrical properties of GaAs. The sheet resistances ($R_s$) of the samples were obtained from van der Pauw measurements. Figure 3a shows the dependence of sheet resistance on various RTA parameters. The sheet resistance drops dramatically from ~$2.1\times10^4$ to 79 Ω/sq as the energy deposited on the sample surface by RTA increases. The decrease in sheet resistance means more dopants have diffused into the GaAs lattice and are electrically activated. The nonmolten GaAs surface remains ~1 nanometer flat when the samples are treated with 700°C for 5 min, as shown in the inset of Figure 3a.

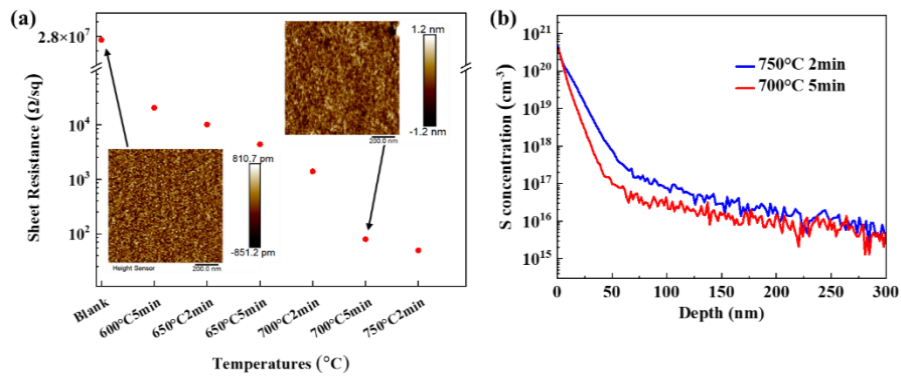

Figure 3. (a) Sheet resistance of S-passivated GaAs with different annealing temperature (insets are the atomic force microscopy images of the sample surfaces). (b) SIMS profile of S in GaAs.

To probe the dopant profile after the RTA treatment (samples S1 and S2 ), we employed secondary ion mass spectrometry (SIMS). Samples S1 and S2 were annealed using 700°C 5min and 750°C 2min, respectively. Among all samples, these two samples have the lowest sheet resistance. Both samples have a comparable dopant profile with a peak concentration of $4\times10^{20}$ cm$^{-3}$ at the surface, as shown in Figure 3b. The sulfur concentration declines rapidly from $4\times10^{20}$ cm$^{-3}$ $10^{17}$ cm$^{-3}$ within 50 nm.

These dopants are carried by the SAMM, which restricts the total number of initial dopants that can be introduced onto the surface. Following high-temperature annealing, the dopants will follow the limited-source diffusion process, which is described by the equation below

$$N(x,t) = \left(\frac{N_0}{\sqrt{\pi Dt}}\right) exp\left[-\left(\frac{x}{2\sqrt{Dt}}\right)^2\right] \quad (1)$$

where $N_0$ represents the initial surface concentration, x denotes the diffusion distance, t is the annealing time, and D is the diffusion constant, which varies with temperature. Based on the SIMS data presented in Figure 3b, we can determine the total amount of dopant atoms that diffuse from the surface into the bulk. Fitting equation (1), we find the average sulfur dopant concentration per unit area is $4.79 \times 10^{14}$ cm$^{-2}$ for the samples annealed at 700°C for 5 minutes, which slightly smaller than the results that Xia et al reported ($6.9 \times 10^{14}$ S atoms·cm$^{-2}$)[19]. One possible reason for this discrepancy is that some sulfur atoms on the GaAs surface may diffuse into the SiO$_2$ capping layer during annealing. This diffusion would lead to a lower sulfur dopant concentration in the GaAs bulk compared to the initial surface coverage.

To get the electrical activation of the dopant atoms, we characterized the charge carrier transport at low temperature when a tunable magnetic field was applied perpendicular to the sample surface. Figure 4a shows the hall resistance at 300K. The resistance decreases linearly monotonically when the applied magnetic field intensity increases in both directions.

Using equation (2), we can calculate the areal concentration of electrons from the derivative of Hall resistance with respect to magnetic field intensity.

$$N_e = -\frac{\Delta B}{e \times (\Delta V_H / I)} \quad (2)$$

where $e$ is the elementary charge, $V_H$ is the Hall voltage, $I$ is the source current, $B$ represents the magnetic field, and $N_e$ is the free electron concentration per unit area. The slope of the linear correlation observed in the Hall measurement gives us the free electron concentration per unit area (denoted as $n_c$) at 300K is extracted to be $3.72 \times 10^{14}$ cm$^{-2}$. Combining the Hall measurements with SIMS results, we can calculate the electrical activation rate of sulfur dopants in GaAs. For samples annealed at 700°C for 5 minutes, the sulfur activation rate is found to be 77.6%. This implies that many sulfur atoms diffuse into the material and become electrically active, while the rest form electrically non-active S-S dimers that do not contribute to the doping effect. For samples annealed at 650°C for 2 minutes and 750°C for 2 minutes (as shown in supplementary material Figure S1), the free electron concentration per unit area is $2.16 \times 10^{11}$ cm$^{-2}$ and $4.85 \times 10^{14}$ cm$^{-2}$, respectively. The increases in the free electron concentration are attributed to longer annealing time and higher annealing temperature. These findings suggest that upon diffusion into the surface, the S atoms may occupy disordered and interstitial sites, necessitating thermal energy to surmount kinetic barriers and facilitate their exchange into substitutional sites, which is crucial for electrical activation[20-22].

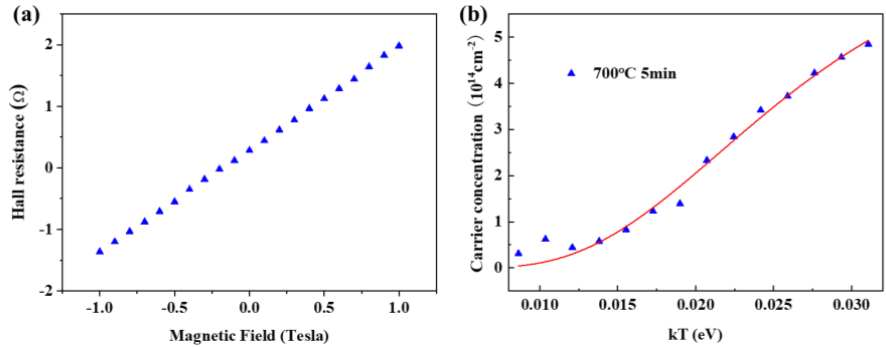

Figure 4. (a) Hall resistance of S-treated GaAs annealed at 700°C for 5 minutes, measured by Hall measurement at room temperature. (b) Temperature dependence of the average free electron concentration for S-treated GaAs.

To further investigate the characteristics of sulfur dopants, we conducted low-temperature Hall measurements to determine their activation energy. We first derived the average free electron concentrations at different temperatures by analyzing the Hall resistance as a function of magnetic field (see supplementary material Table S1 and Figure S2). This analysis allowed us to estimate the free electron concentration in the material at varying temperatures. Next, we plotted the surface concentration $n_c$ as a function of kT (where k is the Boltzmann constant and T is the absolute temperature) in Figure 4b. The dopant distribution in the thin-doped layer near the surface is highly non-uniform, yet it is important to note that the average concentrations of electrons, holes, and ionized dopants will always maintain charge neutrality. Assuming the concentration of holes is negligible compared to that of electrons, we can derive the electron concentration, which varies with temperature and can be expressed using the following mathematical relationship[23].

$$n_c = \frac{-N_c + \sqrt{N_c^2 + 8N_c N_D \exp\left(\frac{\Delta E}{kT}\right)}}{4\exp\left(\frac{\Delta E}{kT}\right)} \quad (3)$$

where $N_c$ is the effective density of states, defined as $N_c \approx 2\left(\frac{2\pi m_n^* kT}{h^2}\right)^{\frac{3}{2}} = w(kT)^{3/2}$ with $w$ being a constant related to the band structure of the semiconductor, $N_D$ represents the donor concentration, and $\Delta E$ is the activation energy, defined as $(E_c - E_d)$. (where $E_c$ is the conductance band edge and $E_d$ is the donor energy level). By fitting equation (3) into the curve in Fig. 4b for the S-doped sample, we obtain that the activation energy of the sulfur dopants as 68 meV and concentration of electrically active S donors as $4.4 \times 10^{14}$ cm$^{-2}$, which aligns well with previous literature[21]. As a further step, we fabricated a p-n junction on p-type GaAs (doped with Zn). We then measured the current-voltage characteristics of the p-n junction and p-type GaAs, as shown in Figures 5. The device has a rectification ratio of up to $10^4$ at ±0.6V bias, the ideality factor of the p-n junction was calculated to be 1.26, indicating the high quality of the fabricated junction.

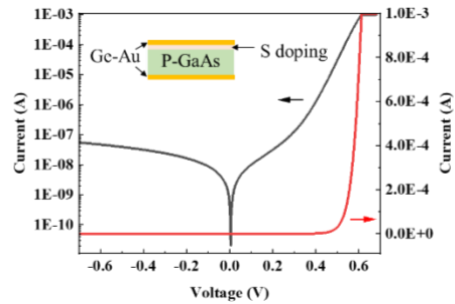

Figure 5. Current–voltage characteristics of S doped p-GaAs (inset is the diagram of the device structure).

Conclusion

In conclusion, we successfully grafted a sulfur monolayer onto the surface of GaAs. Following a 5-minute annealing process at 700°C, the sheet resistance of the sulfur-grafted GaAs samples was reduced to 79 Ω/sq. Through the analysis of SIMS and Hall test results, we determined the electrical activation rate of sulfur dopants in GaAs to be 77.6%, and the donor energy level of sulfur-doped GaAs is located 68 meV below the conduction band. With this process, we successfully fabricated a p-n junction on p-type GaAs (doped with Zn).

Supplementary Material

See the supplementary material that includes the temperature dependance hall resistance and free carrier concentration of S-doped GaAs, a vital physical quantity that characterizes the donor energy level of S-doped GaAs.

Acknowledgments

This work was financially supported by the National Science Foundation of China (NSFC) (No. 62304131 and No. 92065103), the Oceanic Interdisciplinary Program of Shanghai Jiao Tong University (No. SL2022ZD107), the Shanghai Jiao Tong University Scientific and Technological Innovation Funds (No. 2020QY05), and the Shanghai Pujiang Program (No. 22PJ1408200). The devices were fabricated at the

Reference

1  B. A. S. Dunwei Wang, and James R. Heath, Nano Lett **6,** 1096 (2006).
2  J. C. Ho, R. Yerushalmi, Z. A. Jacobson, Z. Fan, R. L. Alley, and A. Javey, Nature Materials **7,** 62 (2007).
3  S. Chang, J. He, S. Prucnal, J. Zhang, J. Zhang, S. Zhou, M. Helm, and Y. Dan, ACS Applied Materials & Interfaces **14,** 30000 (2022).
4  K. Li, J.-Y. Zhang, S. Chang, H. Wei, J.-J. Zhang, and Y. Dan, ACS Applied Electronic Materials **3,** 3346 (2021).
5  P. Taheri, H. M. Fahad, M. Tosun, M. Hettick, D. Kiriya, K. Chen, and A. Javey, ACS Applied Materials & Interfaces **9,** 20648 (2017).
6  N. Li, E. S. Harmon, J. Hyland, D. B. Salzman, T. P. Ma, Y. Xuan, and P. D. Ye, Applied Physics Letters **92** (2008).
7  I. A. Prudaev, S. N. Vainshtein, M. G. Verkholetov, V. L. Oleinik, and V. V. Kopyev, IEEE Transactions on Electron Devices **68,** 57 (2021).
8  R. Yan, D. Gargas, and P. Yang, Nature Photonics **3,** 569 (2009).
9  J. C. Ho, A. C. Ford, Y.-L. Chueh, P. W. Leu, O. Ergen, K. Takei, G. Smith, P. Majhi, J. Bennett, and A. Javey, Applied Physics Letters **95** (2009).
10 N. Tajik, A. C. E. Chia, and R. R. LaPierre, Applied Physics Letters **100** (2012).
11 G. Y. Gu, E. A. Ogryzlo, P. C. Wong, M. Y. Zhou, and K. A. R. Mitchell, Journal of Applied Physics **72,** 762 (1992).
12 V. Richard D'Costa, S. Subramanian, D. Li, S. Wicaksono, S. Fatt Yoon, E. Soon Tok, and Y.-C. Yeo, Applied Physics Letters **104** (2014).
13 Z. H. Lu, M. J. Graham, X. H. Feng, and B. X. Yang, Applied Physics Letters **62,** 2932 (1993).
14 M. S. Carpenter, M. R. Melloch, B. A. Cowans, Z. Dardas, and W. N. Delgass, Journal of Vacuum Science & Technology B: Microelectronics Processing and Phenomena **7,** 845 (1989).
15 M. Hedieh, A. Salehi, and V. R. Mastelaro, Russian Journal of Electrochemistry **57,** 471 (2021).
16 E. C. Mattson and Y. J. Chabal, The Journal of Physical Chemistry C **122,** 8414 (2018).
17 K. R. Kort, P. Y. Hung, P. D. Lysaght, W.-Y. Loh, G. Bersuker, and S. Banerjee, Physical Chemistry Chemical Physics **16** (2014).
18 M. J. Jackson, B. L. Jackson, and M. S. Goorsky, Journal of Applied Physics **110** (2011).
19 W. N. L. H. Xia, G.R. Massoumi , j,j,j, van Eck , L.J. Huang W.M. Lau , D. Landheer, Surface Science **324,** 159 (1995).
20 J. Duan, M. Wang, L. Vines, R. Böttger, M. Helm, Y. J. Zeng, S. Zhou, and S. Prucnal, physica status solidi (a) **216** (2018).
21 E. F. Schubert, in *Doping in III-V Semiconductors* (E. Fred Schubert., 2015), p. 165.
22 S. K. G. R. VENKATASUBRAMANIAN, Journal of Crystal Growth **97** 827—832 (1989).
23 B. Guan, H. Siampour, Z. Fan, S. Wang, X. Y. Kong, A. Mesli, J. Zhang, and Y. Dan, Scientific

Reports **5** (2015).